\documentclass[superscriptaddress,twocolumn,preprintnumbers,amsmath,amssymb,aps,prl]{revtex4-2}

\usepackage{graphicx}
\usepackage{dcolumn}
\usepackage{bm}
\usepackage{hyperref}
\usepackage[mathlines]{lineno}
\usepackage{color}
\usepackage[normalem]{ulem}
\usepackage{rotating}
\def\he3{$^3$He}
\def\Am1{\AA$^{-1}$}
\begin{document}


\title{Momentum distribution of \he3 in one dimension}

\author {M. Boninsegni}
\email{m.boninsegni@ualberta.ca}
\affiliation{Department of Physics, University of Alberta, Edmonton, Alberta, Canada T6G 2H5}

\date{\today}

\begin{abstract}
The one-particle density matrix of a one-dimensional system of fermions featuring a hard-core repulsive interaction at short distances can be computed (numerically) {\em exactly} by means of the continuous-space Worm Algorithm, without any sign instability. We present here results for this quantity, and the related momentum distribution, for a \he3 fluid. It is shown that effects of quantum statistics are observable in  the fully polarized system, but are suppressed in the unpolarized one, atoms being essentially distinguishable in the latter case.
\end{abstract}
\maketitle
Since the seminal work of Girardeau \cite{Girardeau1960}, it has been known that the equilibrium thermodynamic properties of assemblies of identical {\em impenetrable} particles confined to one dimension (1D) are almost entirely independent of quantum statistics. Distinct signatures thereof can be detected only in the momentum distribution $n({q})$, which is experimentally accessible via neutron scattering \cite{Sears1985,Azuah1997,Bryan2016}.
\\ \indent 
Recent advances in the experimental confinement of helium to quasi one dimension \cite{DelMaestro2022,Adams2022} make it at least in principle feasible to measure and compare $n({q})$ for the two isotopes of helium (\he3 and $^4$He), which obey different quantum statistics. This would provide a cogent test of the current theoretical understanding and of the most advanced theories of quantum many-body systems in 1D. There are of course many other quasi one-dimensional Fermi systems (e.g., electronic), for which $n({q})$ is a relevant physical observable. Thus, its reliable theoretical calculation is a worthwhile goal. 
\\ \indent 
The momentum distribution is related through a Fourier transformation to the one-particle density matrix. For assemblies of identical particles in continuous space obeying Bose statistics, this quantity can be generally computed numerically, unbiasedly and to a high degree of precision, using the continuous-space Worm Algorithm (WA) \cite{Boninsegni2006,Boninsegni2006b}, a finite-temperature Quantum Monte Carlo (QMC) method. The application of the WA (or  any other QMC method) to Fermi systems, on the other hand, is hampered by the well-known {\em sign} problem. Thus, although extensive QMC studies of the most abundant (boson)  $^4$He isotope have been carried out, including in (quasi) 1D \cite{Boninsegni2000,Gordillo2000,Boninsegni2001,DelMaestro2011,Bertaina2016,Nava2022},  relatively little has been done for \he3 \cite{Astrakharchic2014} and, to our knowledge, nothing on the momentum distribution.
\begin{figure}
\includegraphics[width=\linewidth]{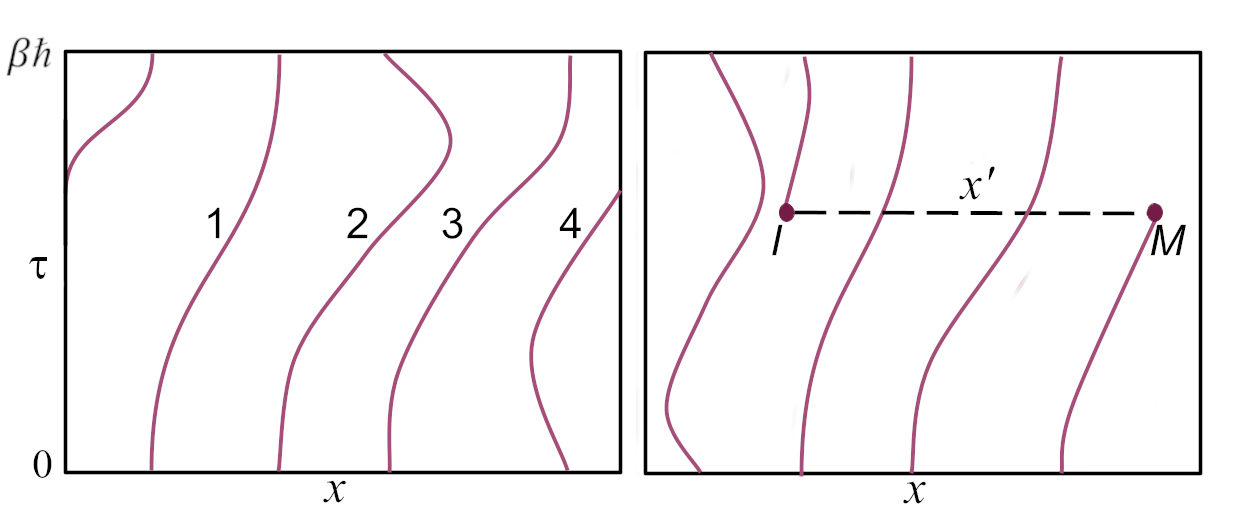}
\caption{{\em Left}: Imaginary time paths (world lines) of four identical particles moving in one dimension. The system has periodic boundary conditions. The temperature is $T=1/\beta$. {\em Right}: Typical configuration generated by the WA contributing to a ``measurement'' of $g(x^\prime)$. In this case the contribution is positive, as there are two closed world lines between Ira (I) and Masha (M).}
\label{worldlines}
\end{figure}
\\ \indent
The origin of the sign problem is readily understood. We focus for definiteness our discussion on the one-particle density matrix
 (but the same basic argument applies to any other physical quantity), defined as
\begin{equation}\label{eqgr}
g({\bf r},{\bf r}^\prime) = \frac{\langle\hat\Psi^\dagger({\bf r}),\hat\Psi({\bf r}^\prime)\rangle}{\langle S\rangle}
 \end{equation}
where $\hat\Psi$, $\hat\Psi^\dagger$ are field operators (either Bose or Fermi), $\langle...\rangle$ stands for thermal expectation value and $S\equiv \eta^P$, with $\eta=(+1)$ ($-1$) for Bose (Fermi) statistics and $P$ is the parity of a permutation of the $N$ identical particles in the system). \\ \indent 
Permutations (and therefore quantum statistics) can be neglected at high temperature, but become increasingly frequent in the low temperature limit, which is when the most remarkable effects of quantum statistics (e.g., Bose condensation) arise. While for Bose systems the denominator of (\ref{eqgr}) is equal to 1, for fermions the contribution from permutations of opposite sign causes both the numerator and the denominator of (\ref{eqgr}) to approach zero \cite{Feynman1965}, in turn making the cost of accurately estimating (\ref{eqgr}) by Monte Carlo grow  {\em exponentially} with $N$. Consequently, obtaining results for systems of size large enough to allow for a reliable extrapolation to the thermodynamic limit proves unfeasible in practice. 
There is presently no systematic, general workaround for this problem, as a result of which  all presently known fermion QMC methods (with few notable exceptions \cite{Dornheim2021}) rely on (to some extent uncontrolled) approximations \cite{Boninsegni1992,Holzmann2011,Militzer2019,Dornheim2019}.
\\ \indent
In 1D and with hard core repulsive interactions, however, 
the {\em only} possible permutations are cyclic, involving \textit{all} $N$ particles in the system, 
This can be visually understood based on Feynman's path integral formulation of quantum statistical mechanics \cite{Feynman1948}. 
Fig. \ref{worldlines} (left part) represents a typical configuration of a system of four identical particles moving in one dimension. Each particle is represented by a world line (path) in imaginary time $0< \tau < \beta\hbar$, where $\beta=1/(k_BT)$, $T$ being the temperature. Paths must be $\beta$-periodic, i.e., the same four positions must be occupied at $\tau=0$ and $\tau=\beta\hbar$, except for a permutation of the particles (e.g., the 4-particle permutation shown in the left part of the figure). Because of the hard core repulsion, paths cannot cross; thus, in 1D a permutation can only occur if one or more paths wind around the boundary, as shown on the right part of Fig. \ref{worldlines}.\\ \indent
This kind of permutation is a finite-size effect, i.e., its contribution to the partition function vanishes in the thermodynamic limit (this statement can be made rigorously \cite{Haldane1981}).  
In practice it is safe to assume that such permutations will not be observed in a simulation comprising a reasonably large number of particles (i.e., at least a few tens) at the temperatures typically of interest.
Because permutations are excluded, the denominator of (\ref{eqgr}) is equal to one, just like for a Bose system. Thus, a Monte Carlo evaluation of (\ref{eqgr}) is possible, even if the numerator is not positive-definite (see below).
This peculiarity of 1D systems has been exploited in the past to carry out QMC calculations of the momentum distribution of Fermi systems on a lattice \cite{Pollet2006}, and in the continuum for the fully polarized electron gas in the ground state \cite{Lee2011}, but to our knowledge no calculation has been attempted at finite temperature and/or for a realistic model of \he3.
\\ \indent
The calculation of  (\ref{eqgr}) for a many-particle system in continuous space can be performed straightforwardly using the WA, which extends the space of many-particle configurations to include an ``open'' world line (worm), which can swap with others as described in Ref. \cite{Boninsegni2006b}. This makes it possible to generate a configuration such as that shown in the right part of Fig. \ref{worldlines}, which illustrates a contribution to $g(x^\prime)$, $x^\prime$ being the spatial distance between the two dangling ends (normally referred to as ``Ira'' and ``Masha''), when they occur at the same $\tau$. For a Fermi system, the sign of the contribution to $g$ is negative (positive) if there is an odd (even) number of closed world lines between Ira and Masha (or, equivalently, one can count the number of ``swap'' operations required to move them apart).  Keeping track of the overall sign simply requires multiplying the running sign by $(-1)$ whenever a swap update is accepted.
\begin{figure}
\includegraphics[width=\linewidth]{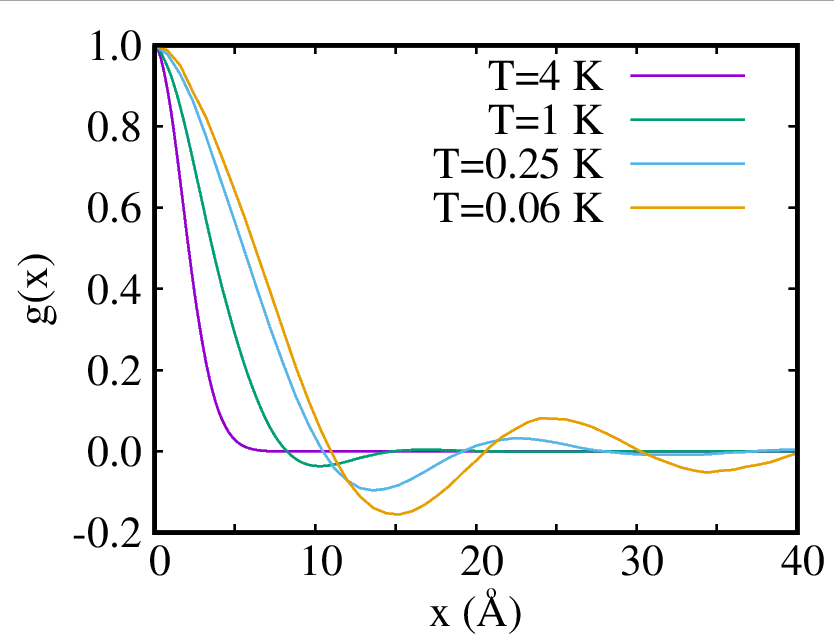}
\caption{ One-particle density matrix of one-dimensional \he3 at different temperatures computed by simulation. The \he3 density is $n=0.1$ \Am1. The system is fully polarized, i.e., the spin projection of all \he3 atoms is the same.}
\label{gofrT}
\end{figure}
\\ \indent
We describe here results obtained for the one-particle density matrix $g(x)$ for one-dimensional \he3, using the WA. The standard microscopic model of helium is used here, based on the well known Aziz pair potential \cite{Aziz1979}. The system includes $N$ particles enclosed in a box of length $L$ (i.e., the linear density is $n=N/L$) with periodic boundary conditions. Details of the simulation are standard (the reader is referred to Ref. \cite{Boninsegni2006b} for details). The momentum distribution is obtained as the Fourier transform of the quantity $ng(x)$. 
\\ \indent
The number of particles utilized was typically chosen to keep the product $LT$ constant, for a given density. The simulation carried out at the lowest temperature, namely $T=0.0625$ K ($T=0.125$ K) for $n=0.1$ \Am1 ($n=0.2$ \Am1) included $N=64$ ($N=96$) particles. For $n=0.3$ \Am1 the results are found to be essentially temperature and size independent, within the precision of our calculation, for $T < 1$ K. 
For all densities, we performed simulations at the lowest temperature on a system of size twice as large as well, as a check, obtaining results consistent with those obtained for a smaller system size.
\begin{figure}[t]
\includegraphics[width=\linewidth]{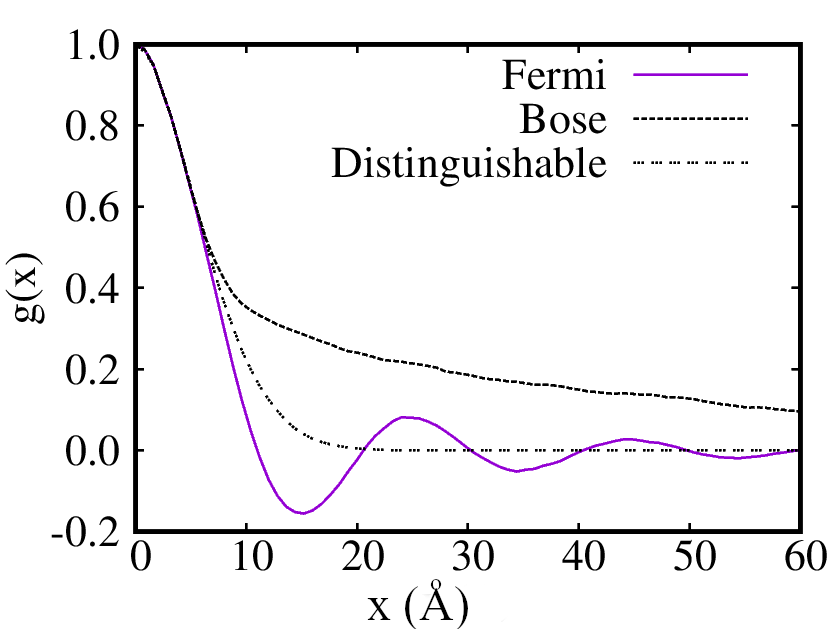}
\includegraphics[width=\linewidth]{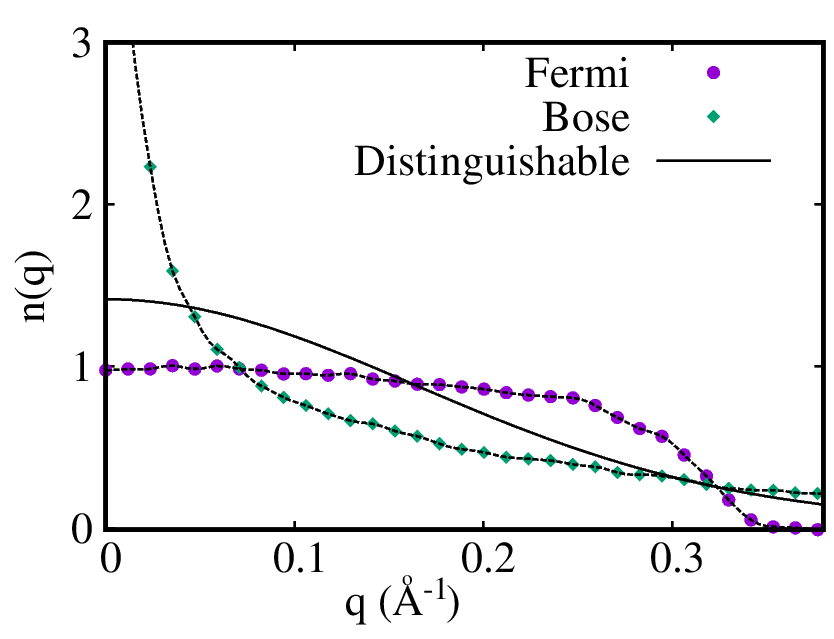}
\caption{{\em Top:} One-particle density matrix of one-dimensional \he3 at density $n=0.1$ \AA$^{-1}$ and temperature $T=0.06$ K, computed by simulation. The system is fully polarized, i.e., the spin projection of all \he3 atoms is the same. Also shown for comparison is the same quantity computed for a hypothetical Bose \he3 system (dotted line) and for distinguishable particles (dashed line). The simulated system comprises $N=64$ atoms. {\em Bottom}: Corresponding momentum distribution $n(q)$ for the three cases. Statistical and systematic errors are estimated to be of the order of the size of the symbols. }
\label{gofrnq}
\end{figure}
\\ \indent
A preliminary remark is that since \he3 atoms are spin-1/2 fermions there are two possible spin projection states (we shall henceforth refer to them as $u$ and $d$, with an obvious notation). Because of the hard core repulsion, the atomic arrangement (e.g., \textit{uududdu...} from left to right) cannot change in the course of a simulation. Since the swap move can only be attempted between two world lines corresponding to identical particles, the effect of quantum statistics is most important if all atoms have the same (either) spin projection, i.e., if the system is \textit{fully polarized}; it is progressively reduced if the polarization is only partial \cite{notx}.
\\ \indent
We begin with considering a fully polarized system (the unpolarized case is discussed below). Fig. \ref{gofrT} shows the $g(x)$ for a system of density $n=0.1$ \Am1, computed at different temperatures, the highest being 4 K, the lowest 60 mK. At high temperature effects of quantum statistics are negligible and the one-body density matrix approaches a Gaussian function. As the temperature is lowered, however, it develops the characteristic oscillation of period equal to twice the average interatomic distance, which is a distinct feature of Fermi systems. As expected, the statistical error does not grow significantly as the temperature is lowered; it remains unnoticeable on the scale of the figure, confirming the absence of any QMC sign problem for this particular fermion problem.
\begin{figure}
\includegraphics[width=\linewidth]{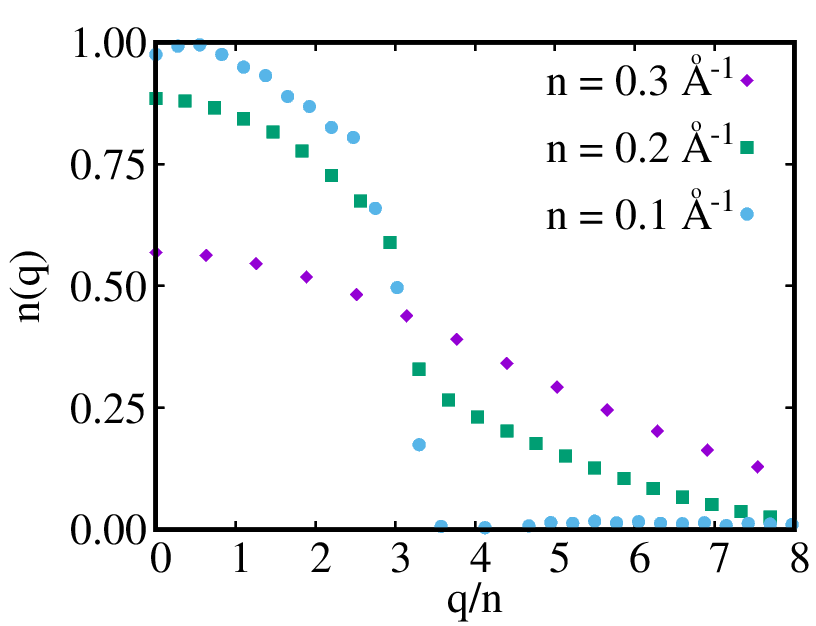}
\caption{{\em Top:} Computed momentum distribution of one-dimensional \he3 for different linear densities, at temperature $T=0.08$ $T_F$, the Fermi temperature. In order to facilitate the comparison, the wave vector is expressed in units of density. The system is fully polarized, i.e., the spin projection of all \he3 atoms is the same. Statistical and systematic errors are of the order of the size of the symbols.}
\label{comparenq}
\end{figure}
\\ \indent
In fig. \ref{gofrnq} (top) the $g(x)$ computed at $n=0.1$ \Am1  and $T=0.06$ K is compared with that obtained assuming that atoms are distinguishable, as well as for a  fictitious (spin-zero) Bose  system with the same mass and interaction.  While the three results coincide in the $x\to 0$ limit, the Bose system has a slowly  decaying, positive-definite $g(x)$. The differences are reflected in the corresponding momentum distributions, shown in the bottom part of Fig. \ref{gofrnq}; specifically, for fermion \he3 $n(q)$ approaches the well-known Fermi-Dirac distribution, with a cutoff around the Fermi wave vector, where it is smeared by thermal and interaction effects. In sharp contrast, the Bose system displays a long tail and a peak at zero momentum, despite the absence of Bose condensation in 1D. The result for indistinguishable particles is essentially classical, and in a way interpolates between Fermi and Bose statistics. 
\begin{figure}
\includegraphics[width=\linewidth]{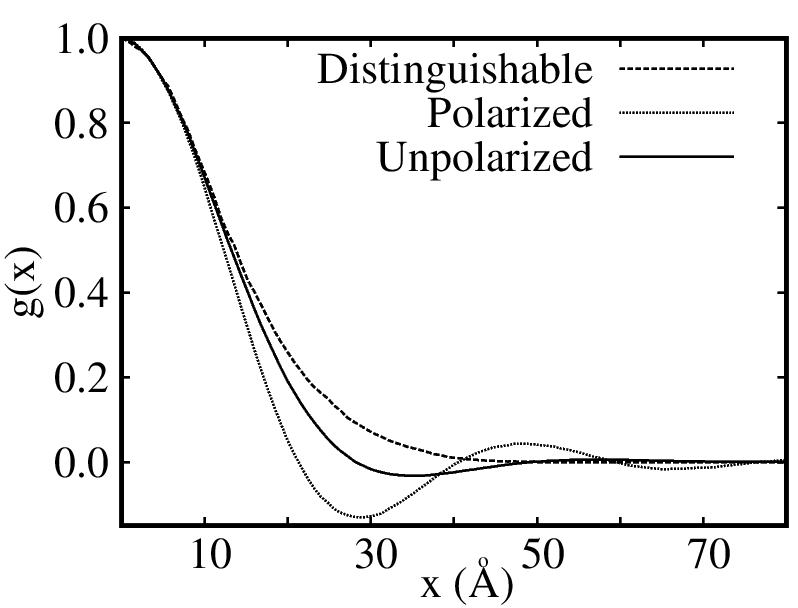}
\includegraphics[width=\linewidth]{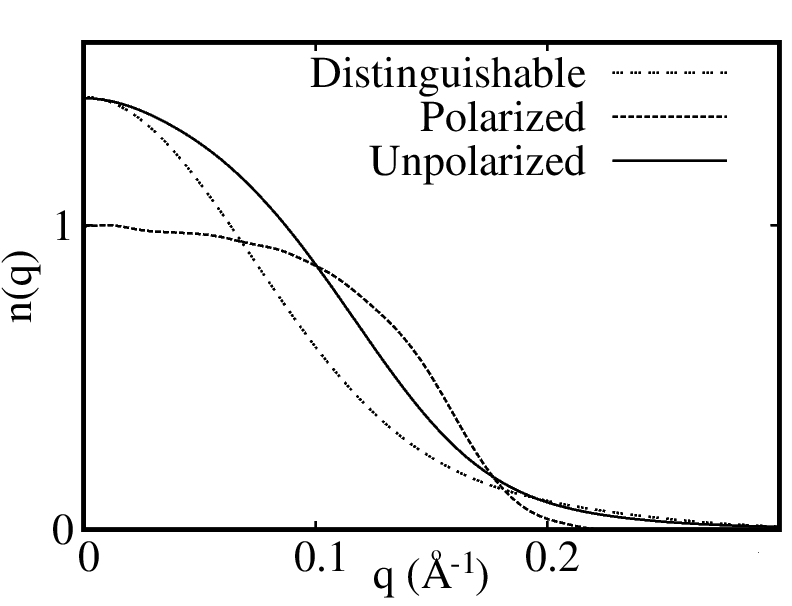}
\caption{{\em Top:} One-particle density matrix of unpolarized one-dimensional \he3 at temperature $T=0.03$ K, computed by simulation for atoms of the same spin projection (solid line). The \he3 density is 0.05 \AA$^{-1}$. Dotted line is for fully polarized \he3, dashed line for distinguishable particles. {\em Bottom}: Corresponding momentum distribution $n(q)$. For comparison purposes, the momentum distribution for the unpolarized case has been rescaled so that its value at $q=0$ is the same as that of a system of distinguishable particles.}
\label{comprq}
\end{figure}
\\ \indent
It is interesting to explore the dependence of the momentum distribution on the density of fermions. Fig. \ref{comparenq} shows results for three different densities at temperature $T=0.08\ T_F$, where $T_F$ is the Fermi temperature. To carry out a meaningful comparison we express the wave vector in units of density and  rescale all curves so that the value at $q=0$ is 1. The result at the lowest density considered here (0.1 \Am1) features the sharpest drop in the vicinity of the Fermi wave vector, whereas at $n=0.2$ \Am1, the smearing is more significant, as interactions play more important a role; however, the general shape of the curve is largely the same in these two cases. 
It is worth noting that, at $n=0.2$ \Am1 the system is in the (quasi) crystalline phase, as can be established by computing the Luttinger parameter (Ref. \cite{Astrakharchic2014}, independently verified here), as well as by the appearance of a sharp peak in the static structure factor at $q=2\pi n$. Yet, effects of quantum statistics are clearly detectable.
\\ \indent 
The conventional wisdom is that for a hard core system effects of quantum statistics are suppressed in the crystalline phase; indeed, it seems now accepted that the occurrence of a ``supersolid'' phase, combining density long-range order with broken gauge symmetry hinges on the softness of the repulsive part of the interaction at short distances \cite{Boninsegni2012c,Kora2019}. However, atomic exchanges are known to be important in the bulk solid phase of $^3$He (see, for instance, Ref. \cite{Landesman2017}). It may therefore be regarded as unsurprising that this should be {\em a fortiori} true in 1D.
At sufficiently high density, on the other hand, the momentum distribution changes qualitatively, turning into a Gaussian, as the one-particle density matrix $g(x)$ is identical (within the precision of our simulation) for all three types of statistics (Bose, Fermi, distinguishable particles). This is shown in Fig. \ref{comparenq} for the highest density considered here, namely $n=0.3$ \Am1.
\\ \indent
Next we discuss the results for \textit{unpolarized} \he3. As mentioned in the introduction, the presence of  a hard core repulsion renders it impossible for atoms to trade place with their adjacent neighbors. As a result, the initial alignment of atoms of different spin projections is unaltered in the course of the simulation. The one-particle density matrix will therefore in general depend on the specific arrangement of particles \cite{notx}. In order to obtain a result as close as possible to realistic experimental conditions, we have computed the $g(x)$ by averaging over a number (typically 20) of different random arrangements of $u$ and $d$ \he3 atoms. 
\\ \indent
Fig. \ref{comprq} shows the results for $g(x)$ and the ensuing $n(q)$ for a system of density $n=0.05$ \Am1, at temperature $T=0.03$ K. As expected, the one-body density matrix for same-spin atoms (top) features much less pronounced oscillations compared to the polarized case (shown in the figure by the dotted line). As explained above, this arises in the calculation from the impossibility of ``swapping'' adjacent world lines pertaining to different types of particles (i.e., $u$ and $d$). \\ \indent
While some weak signatures of quantum statistics remain (mainly the dip in the $g(x)$, that goes slightly negative at $x\sim 30$ \AA), in the absence of polarization the system essentially behaves as if atoms were distinguishable. This is manifest in the momentum distribution,  shown in the lower  panel of Fig. \ref{comprq}. The jump in correspondence of the Fermi momentum (typically associated with the quasiparticle spectral weight) is washed out for the unpolarized system, and the curve is altogether qualitatively more similar to that for a system of distinguishable particles (dashed line), e.g., the presence of a tail at large $q$. It is important to notice that this is observed at a relatively low density, and that that an even stronger suppression of the effect of quantum statistics can be expected at higher density.
\\ \indent
In summary, we outline a procedure allowing the calculation by QMC of the one-particle density matrix of a system of hard core fermions in one dimension,  without any approximation and any sign instability. The computational scheme is a straightforward extension of the continuous-space WA. Although the results provided here are for \he3, the same algorithm can be applied to any system of fermions in one dimension, provided that there be a hard core repulsion at short distances. \\ \indent 
It is known that effects of quantum statistics are often observable in physical settings in which exchanges of indistinguishable particles are strongly suppressed, e.g., in liquid parahydrogen near freezing \cite{Boninsegni2009}. It is shown that, for  systems of impenetrable particles in 1D, whose equilibrium thermodynamic properties are largely independent of quantum statistics, clear signatures thereof can be detected in the momentum distribution, as long as the system is spin-polarized.
\\ \indent
This work was supported by the Natural Sciences and Engineering Research Council of Canada. The author gratefully acknowledges valuable discussions with N. V. Prokof'ev.

\bibliography{references}
\end{document}